\documentclass[prb,preprint]{revtex4-1} 
\usepackage[table]{xcolor}

\newcolumntype{s}{>{\columncolor[HTML]{AAACED}} p{3cm}}

\usepackage{amssymb}
\usepackage{epsfig,amsmath,graphics}
\usepackage{epstopdf}
\usepackage{graphicx}
\usepackage{bm}
\usepackage{gensymb}
\usepackage{array}
\usepackage{multirow}

\ifdefined\WORDCOUNT
    \newcommand{\HideEq}[1]{}
\else
    \newcommand{\HideEq}[1]{#1}
\fi

\def\be{\begin{equation}}
\def\ee{\end{equation}}

\begin{document}

\title{Physically significant phase shifts in matter-wave interferometry}

\author{Chris Overstreet}
\email{coverstr@stanford.edu}

\author{Peter Asenbaum}

\author{Mark A. Kasevich}

\affiliation{Department of Physics, Stanford University, Stanford, California 94305}

\date{October 22, 2020}

\begin{abstract}
Many different formalisms exist for computing the phase of a matter-wave interferometer.  However, it can be challenging to develop physical intuition about what a particular interferometer is actually measuring or about whether a given classical measurement provides equivalent information.  Here we investigate the physical content of the interferometer phase through a series of thought experiments.  In low-order potentials, a matter-wave interferometer with a single internal state provides the same information as a sum of position measurements of a classical test object.  In high-order potentials, the interferometer phase becomes decoupled from the motion of the interferometer arms, and the phase contains information that cannot be obtained by any set of position measurements on the interferometer trajectory.  This phase shift in a high-order potential fundamentally distinguishes matter-wave interferometers from classical measuring devices.

\end{abstract}
\maketitle

\section{Introduction}

Matter-wave interference is the quintessential quantum-mechanical phenomenon.  Ever since the publication of the Feynman Lectures\cite{Feynman1965} in 1965, students have been introduced to the principles of quantum mechanics through the example of double-slit interference.  Modern matter-wave interferometers, which employ multiple beamsplitters and mirrors, have similarities with optical interferometers (Fig. 1).  In both cases, an incident wave is coherently split into two trajectories by an initial beamsplitter.  The trajectories are reflected toward one another and interfered on a final beamsplitter, leading to a phase-dependent intensity in each of the two outgoing trajectories (``output ports'').  

\begin{figure}[h]
	\begin{center}
		\includegraphics[width=\columnwidth]{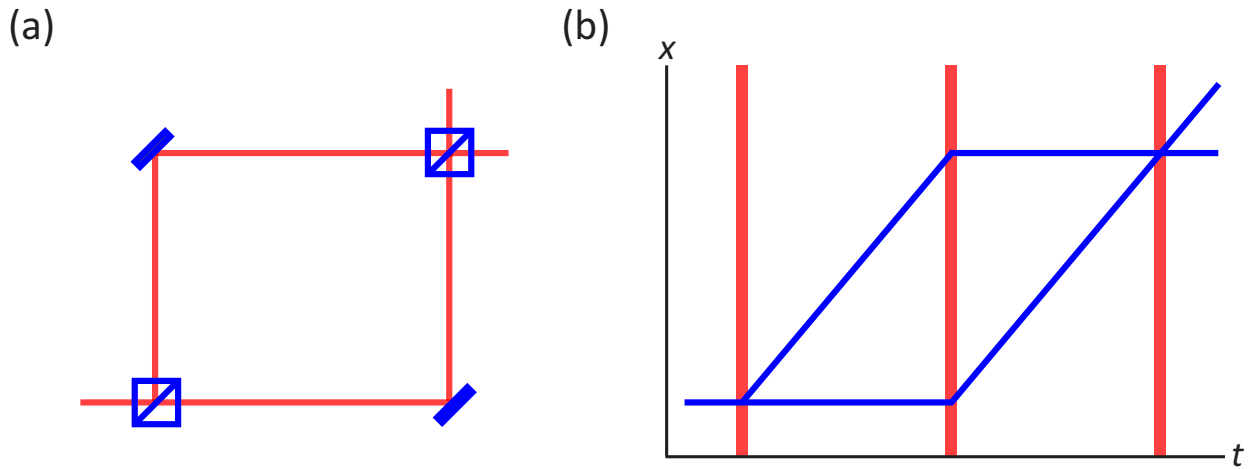}
        \caption{(a) Optical interferometer.  A laser (red) is split onto two trajectories that are reflected and interfered by optical elements (blue). (b) Spacetime diagram of a matter-wave interferometer.  A matter wave (blue) is split onto two trajectories that are reflected and interfered by matter-wave beamsplitters and mirrors (red), which may consist of light pulses or material gratings.}
	\end{center}
\end{figure}

In addition to their central role in demonstrating the validity of quantum mechanics,\cite{Schrinski2020} matter-wave interferometers are useful measuring devices with a variety of scientific applications.  Atom interferometers have been used to test the equivalence principle of general relativity at the $10^{-12}$ level,\cite{Asenbaum2020} to measure the fine-structure constant with an accuracy of $0.2$~ppb,\cite{Parker2018} and to produce accurate gyroscopic\cite{Gustavson2000,Stockton2011} and gravity-gradiometric\cite{McGuirk2002,Asenbaum2017} measurements.  Proposed future experiments based on atom interferometry include km-scale underground and space-based equivalence principle tests\cite{Aguilera2014} and gravitational wave detectors.\cite{Canuel2018,Coleman2018}  

The result of an interferometric measurement consists of the interferometer phase, which is determined by the number of particles in each output port.  The interferometer phase is equal to the phase difference between interferometer arms.  In an optical interferometer, the phase is simply related to the optical path length of each arm.\cite{Born1999}  Calculating the phase of a matter-wave interferometer is more complicated than in the optical case because of nontrivial contributions from the action difference$^{13}$
between arms.  Several different approaches have been developed to compute the phase, including the midpoint theorem,\cite{Antoine2003,Antoine2006} the Wigner function method,\cite{Dubetsky2006,Giese2014} a representation-free description,\cite{Kleinert2015,Bertoldi2019} and techniques that solve the Schr\"odinger equation\cite{Hogan2009,Dimopoulos2008,Roura2020} in various approximations.  Due to the diverging formalisms, it can be difficult to form intuition about the physical content of a given interferometric measurement.

Physical intuition about any experiment can be developed in the following way:  by using the simplest possible formalism to describe a given experimental situation, by focusing on physical observables rather than calculational artifacts, and by comparing the experiment to other measuring devices.  In this article, we will attempt to use this approach to build physical intuition for matter-wave interferometry.  While our discussion will focus on light-pulse atom interferometry\cite{Hogan2009} as a particular example, our analysis (and the associated physical intuition) is valid for a broad class of matter-wave interferometers, including neutron interferometers,\cite{Colella1975} atom interferometers with material gratings,\cite{Keith1991} and guided interferometers.\cite{Shin2004,Wang2005,Xu2019}       

The remainder of the paper is organized as follows:  Section \ref{Sec:classicalAccel} discusses a classical accelerometer; Section \ref{Sec:AIAccel} computes the phase of a matter-wave accelerometer from the midpoint theorem and describes the analogy between classical and interferometric measurements; Section \ref{Sec:BeyondMP} explains the semiclassical method for computing the full interferometer phase, modifying the usual formalism to highlight its physical meaning; Section \ref{Sec:PropToMP} contains a derivation of the relationship between the semiclassical method and the midpoint theorem; and Section \ref{Sec:Discussion} summarizes the physical intuition obtained.  

\section{A classical accelerometer}

\label{Sec:classicalAccel}

Suppose we want to measure our acceleration with respect to a freely falling reference frame.  One way to do this is to observe the motion of a freely falling object, recording the position of the object at various times.  If the object is located at position $x_0$ and has velocity $v_0$ at time $t = 0$, then its position $x(t)$ at time $t$ is 
\be
x(t) = x_0 + v_0 t - \frac{1}{2} g t^2
\ee
where $g$ is the relative acceleration.  This equation has three unknowns:  $x_0$, $v_0$, and $g$.  Thus, we must measure the position at three times in order to solve for $g$.  If the total time available for the observation is $2T$, we can choose to make position measurements at times $t = 0$, $t = T$, and $t =  2T$.  Solving for $g$ in terms of these measurements, we have       
\be
g = -\frac{1}{T^2}\left[x(0) -2 x(T) + x(2T)  \right].
\ee
A position measurement may be understood as a dimensionless measurement of the phase $\phi$ of some ruler with wavenumber $k$, i.e. $x = \phi/k$ (Fig. 2).  In terms of $\phi$, we have
\be
g = -\frac{1}{kT^2}\left[\phi(0) -2 \phi(T) + \phi(2T)  \right].
\ee

\begin{figure}[h]
	\begin{center}
		\includegraphics{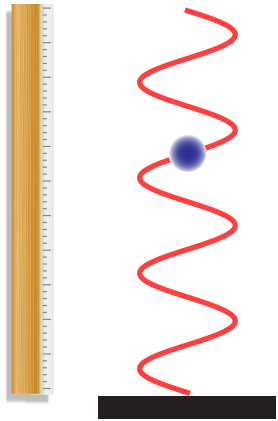}
        \caption{Position measurement.  The displacement between a test object (blue) and reference object (black) is measured by observing the phase of a ruler with a well-defined wavenumber.  An optical standing wave can be used as a ruler.}
	\end{center}
\end{figure}

\section{An atom-interferometric accelerometer}

\label{Sec:AIAccel}

Now suppose that instead of a classical object, we use a freely falling cloud of atoms as an acceleration reference.  Rather than making three position measurements, we implement a Mach-Zehnder light-pulse atom interferometry sequence \cite{Kasevich1992} (Fig. 3a) with atom-light interactions at times $t \in \{0, T, 2T\}$ and measure the resulting interferometer phase.  

\begin{figure}[h]
	\begin{center}
		\includegraphics[width=\columnwidth]{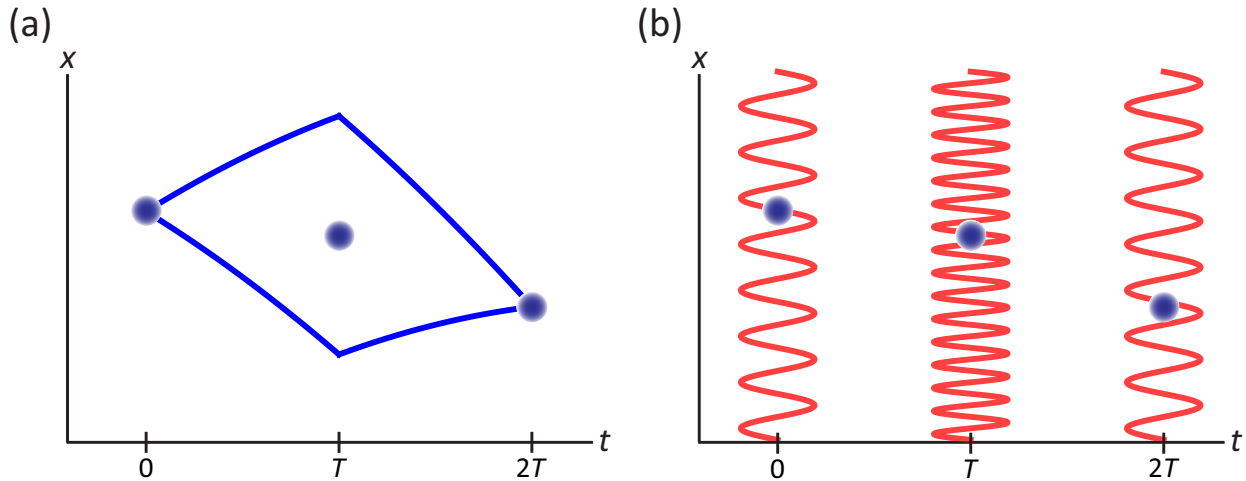}
        \caption{(a) Spacetime diagram of a Mach-Zehnder accelerometer.  The atoms are split, redirected, and interfered by a sequence of three light pulses. Blue points represent the midpoint position of the interferometer during each light pulse.  (b) Mach-Zehnder accelerometer in the midpoint theorem representation.  The interferometer phase depends on the interferometer midpoint positions as well as the wavevectors of the optical gratings.}
	\end{center}
\end{figure}

The interferometer phase in this example can be calculated by using the midpoint theorem.\cite{Antoine2003}  As shown in Fig. 3b, the midpoint theorem represents the interferometer as a sequence of effective atom-light interactions.  To fix notation, we define $x_{j, i}$ to be the displacement of arm $j$ with respect to the laser at the time of the $i^\text{th}$ pulse.  Each atom-light interaction is treated as an effective coupling of two atomic states by a laser of wavevector $k_{i}$, frequency $\omega_{i}$, and phase $\phi_{i}$.  In each atom-light interaction involving arm $j$, the atoms on arm $j$ either absorb one photon (gaining momentum $\hbar k_i$ and internal energy $\hbar \omega_i$) or emit one photon (losing momentum $\hbar k_i$ and internal energy $\hbar \omega_i$).  We can therefore define the wavevector $k_{j,i}$, frequency $\omega_{j,i}$, and phase $\phi_{j,i}$ of the $i^\text{th}$ interaction on arm $j$ as $k_{j,i} = \pm k_i$, $\omega_{j,i} = \pm \omega_i$, and $\phi_{j,i} = \pm \phi_i$, where the sign depends on whether arm $j$ absorbs $(+)$ or emits $(-)$ a photon during the $i^\text{th}$ interaction.  If the $i^\text{th}$ interaction leaves the momentum and internal energy of arm $j$ unchanged, then $k_{j,i} = \omega_{j,i} = \phi_{j,i} = 0$.  Note that in practice, multi-photon transitions (such as Bragg\cite{Martin1988} or Raman\cite{Kasevich1991} transitions) are often used to couple the two atomic states; in that case, the wavevector, frequency, and phase of the effective coupling must be calculated by adiabatic elimination of intermediate states.\cite{Mueller2008}  

Atoms can exit the interferometer in one of two output ports, each of which corresponds to a particular momentum state.  The interferometer phase $\phi$ determines the probabilities $P_1$ and $P_2$ that an atom will be found in a given output port according to the expression
\be
P_1 = \frac{1}{2} + \frac{1}{2} \cos \phi; \; \; \; P_2 = \frac{1}{2} - \frac{1}{2} \cos \phi
\ee
where the output port labels have been chosen so that port 1 corresponds to the input port.  The midpoint theorem states that the interferometer phase is given by
\be \label{eq:MPTheorem}
\phi_\text{MP} \equiv \sum_{i = 1}^N \left[ (k_{1, i} - k_{2, i})\;\bar{x}_i -(\omega_{1,i} - \omega_{2,i})\;t_i + (\phi_{1,i} - \phi_{2,i} \right)].  
\ee
Here the index $i$ runs over the $N$ atom-light interactions, $t_i$ is the time of the $i^\text{th}$ pulse, and
\be
\bar{x}_i \equiv \frac{x_{1,i} + x_{2,i}}{2} 
\ee
is the average displacement of the two arms with respect to the laser at time $t_i$.

To calculate the interferometer phase, we choose one output port and add up the phase shifts due to each interaction between the light and the atoms in that output port.  The interferometer phase does not depend on which output port is used for the calculation.  According to the midpoint theorem, we may write the phase $\phi_\text{MZ}$ of the Mach-Zehnder interferometer as 
\be
\phi_\text{MZ} = \sum_{i = 1}^{3} (k_{1,i} - k_{2,i})\;\bar{x}_i.
\ee
Here we are assuming that $\phi_i$ is constant for all $i$.  We are also using the fact that in a Mach-Zehnder accelerometer, the terms involving $\omega_{i}$ sum to zero because the two arms spend equal time in each internal state. 

Simplifying the sum, we have
\be
\phi_\text{MZ} = k\, \bar{x}(0) - 2 k\, \bar{x}(T) + k\, \bar{x}(2T) = -k g T^2
\ee
or equivalently
\begin{align}
g &= -\frac{1}{k T^2}\left[\phi(0) -2 \phi(T) + \phi(2T)  \right] \\
&= -\frac{1}{T^2}\left[\bar{x}(0) -2 \bar{x}(T) + \bar{x}(2T)  \right]. 
\end{align}

The interferometer phase thus contains precisely the same acceleration information as the three-point classical accelerometer described in the previous section.  We may think of the atom-light interactions as providing three position measurements of the interferometer midpoint with respect to the laser, and the Mach-Zehnder pulse sequence combines these position measurements such that the phase gives the relative acceleration.$^{32}$

From the form of the midpoint theorem, which expresses the interferometer phase as $\phi \sim \sum k_i \bar{x}_i$ up to terms proportional to $\omega_i$, we see that the analogy between a matter-wave interferometer and a classical test mass holds whenever the midpoint theorem is valid.  For any interferometer in which the phase can be described by the midpoint theorem, there exists a corresponding classical experiment in which: (i) the classical test mass travels along the midpoint trajectory of the interferometer, and (ii) the interferometer phase may be written in terms of the positions of the classical test mass at the times of the interferometer beamsplitters and mirrors (up to ``clock'' phase shifts proportional to $\omega_i$).

\section{Beyond the midpoint theorem}

\label{Sec:BeyondMP}

The midpoint theorem is an effective theory rather than a first-principles calculation.  What is the midpoint theorem's range of validity, and how are phase shifts beyond the midpoint theorem computed?    

The most straightforward way to perform a first-principles calculation of the interferometer phase is to represent the interferometer's initial state by a quantum-mechanical wavefunction and solve the time-dependent Schr\"odinger equation.  This technique has been applied to light-pulse atom interferometry in both non-relativistic\cite{Hogan2009} and relativistic\cite{Dimopoulos2008,Roura2020} settings.  The time evolution is divided into intervals of atom-light interaction separated by  time intervals in which the laser intensity is zero (``drift times'').  The $i^\text{th}$ atom-light interaction is assumed to couple two atomic states with a photon of wavevector $k_i$, frequency $\omega_i$, and phase $\phi_i$; between the atom-light interactions, the wavefunction evolves with potential energy $V(\hat{x}, t)$.  

The key assumption that makes the calculation analytically tractable is the \textit{semiclassical approximation}, which stipulates that the higher-order coordinate dependence of the Hamiltonian is negligible at the size scale of the wavepacket on each interferometer arm.  Specifically, if $H(\hat{x}, \hat{p}, t)$ is the Hamiltonian between interferometer pulses (i.e. when there is no atom-light interaction), then we can expand Ehrenfest's theorem around the central position and momentum of each wavepacket to obtain
\be \label{Eq:EhrenfestX}
\partial_t \left< \hat{x} \right> = \partial_{\hat{p}} H + \frac{1}{2!} \partial_{\hat{p}} \partial_{\hat{p}} \partial_{\hat{p}} H \cdot \Delta p^2 + \cdots
\ee
and 
\be \label{Eq:EhrenfestP}
\partial_t \left< \hat{p} \right> = -\partial_{\hat{x}} H - \frac{1}{2!} \partial_{\hat{x}} \partial_{\hat{x}} \partial_{\hat{x}} H \cdot \Delta x^2 + \cdots.
\ee
In these expressions, $\Delta x^2 \equiv \left< \hat{x}^2 \right> - \left< \hat{x} \right>^2$ and $\Delta p^2 \equiv \left< \hat{p}^2 \right> - \left< \hat{p} \right>^2$ are the position and momentum variance of the wavepacket, respectively.  The semiclassical approximation asserts that all of the higher-order terms on the right-hand sides of Eq.~\ref{Eq:EhrenfestX} and Eq.~\ref{Eq:EhrenfestP} are zero.  Physically, the semiclassical approximation implies that phase shifts within the wavepacket due to high-order Hamiltonian terms are small compared to the phase resolution.  Then for each wavepacket, we have $\partial_t \left< \hat{x} \right> = \partial_{\hat{p}} H$ and $\partial_t \left< \hat{p} \right> = -\partial_{\hat{x}} H$.  With the semiclassical approximation, Ehrenfest's theorem reduces to Hamilton's equations for the wavepacket on each interferometer arm, and the expectation values of position and momentum on the $j^\text{th}$ arm follow the classical trajectories:
\be
\left< \hat{x}_j \right>(t) = x_j(t)
\ee
and
\be
\left< \hat{p}_j \right>(t) = p_j(t).
\ee

Notice that the semiclassical approximation requires variations of the Hamiltonian to be small with respect to the size of the wavepacket on each interferometer arm, but the variations need not be small with respect to the position or momentum separations between interferometer arms.  The semiclassical approximation therefore remains valid in long-time, large-momentum-transfer atom interferometry\cite{Kovachy2015a} as long as the wavepackets on each arm are sufficiently narrow in position and momentum.$^{34}$

It can then be shown\cite{Hogan2009} that the interferometer phase $\phi$ is given by
\be
\phi = \phi_\text{laser} + \phi_\text{prop} + \phi_\text{sep}. 
\ee
The first term, known as the ``laser phase,'' is given by 
\be
\phi_\text{laser} = \sum_{i = 1}^{N} \left[ \left( k_{1,i}\, x_1(t_i) - \omega_{1,i}\, t_i + \phi_{1,i} \right)  - \left( k_{2,i}\, x_2(t_i) - \omega_{2,i}\, t_i + \phi_{2,i} \right) \right].
\ee
In this expression, the index $i$ runs over the $N$ atom-light interactions.  The laser wavevectors $k_{j,i}$, frequencies $\omega_{j,i}$, and phases $\phi_{j,i}$ are defined as in Section \ref{Sec:AIAccel}. 

The second term in the interferometer phase, called the ``propagation phase,'' is the difference in the action computed along the classical trajectory of each interferometer arm.  It is given by 
\be
\phi_\text{prop} = \frac{1}{\hbar} \int_{t_0}^{t_f} \mathcal{L}(x_1(t), p_1(t), t)\; dt   - \frac{1}{\hbar} \int_{t_0}^{t_f} \mathcal{L}(x_2(t), p_2(t), t)\; dt  
\ee
where $\mathcal{L}$ is the Lagrangian, $t_0$ is the time of the initial beamsplitter, and $t_f$ is the time of the final beamsplitter.$^{13}$

The third term, the ``separation phase,'' is given by 
\be
\phi_\text{sep} = \frac{1}{\hbar}\frac{p_1(t_f) + p_2(t_f)}{2}[x_2(t_f) - x_1(t_f)].
\ee

Notice that the semiclassical approximation reduces the calculation of the phase from a quantum-mechanical problem to a classical one.  In order to compute the interferometer phase with this approximation, it is sufficient to know the classical trajectory of each interferometer arm, and the dynamics of the wavepacket on each interferometer arm can be ignored.  

Although it provides a significant computational simplification, the semiclassical approximation introduces calculational artifacts that complicate efforts to understand the physical meaning of each term in the interferometer phase.  These artifacts arise because the classical trajectories of the interferometer arms need not spatially overlap at the time of the final beamsplitter pulse.  In such cases, which are called ``open interferometers,''$^{35}$
the terms in the interferometer phase acquire the following properties.  First, the separation phase becomes nonzero.  Second, the propagation phase becomes frame-dependent.  Transforming to a coordinate system that moves at a relative velocity $v'$ adds a term $-m v'\, [x_1(t_f) - x_2(t_f)]/\hbar$ to the propagation phase that is canceled by a term of the opposite sign in the separation phase. Third, the laser phase and separation phase become dependent on which output port is chosen to compute them.  Using the opposite output port, which moves at a relative velocity $\hbar k/m$, adds a term $k\, [x_1(t_f) - x_2(t_f)]$ to the laser phase that is canceled by a term of the opposite sign in the separation phase.

Physical effects are always independent of one's choices of coordinate system and calculational method.  To reveal the physical content of the semiclassical formalism, we can reapproach it in the following way:  rather than calculating the interferometer phase in a single output port, we calculate each phase term in both output ports and average the results.  The interferometer phase is then given by 
\be
\phi = \bar{\phi}_\text{laser} + \bar{\phi}_\text{loop}  
\ee
where $\bar{\phi}_\text{laser}$ is the laser phase averaged between the two ports and $\bar{\phi}_\text{loop} \equiv \phi_\text{prop} + \bar{\phi}_\text{sep}$ is the port-averaged sum of the propagation and separation phase.  We may think of $\bar{\phi}_\text{loop}$ as an integral around the closed interferometer trajectory, with $\bar{\phi}_\text{sep}$ providing the last piece of the closed-loop integral that is missing from  $\phi_\text{prop}$.$^{36}$

This way of arranging the phase terms has several advantages.  First, unlike $\phi_\text{prop}$ and $\phi_\text{sep}$, $\bar{\phi}_\text{loop}$ is invariant under Galilean transformations.  In fact, since it is the action of a closed trajectory, $\bar{\phi}_\text{loop}$ is invariant under point transformations $(x, \dot{x}, t) \rightarrow (y(x, t),\, \dot{y}(x, \dot{x}, t),\, t)$.  Second, unlike $\phi_\text{laser}$ and $\phi_\text{sep}$, $\bar{\phi}_\text{laser}$ and $\bar{\phi}_\text{loop}$ do not depend on the arbitrary choice of which output port to use for the calculation.  

Due to their dependence on calculational artifacts, it is not possible to assign a physical meaning to $\phi_\text{prop}$, $\phi_\text{laser}$, or $\phi_\text{sep}$ individually.  However, it is reasonable to offer a physical interpretation of $\bar{\phi}_\text{laser}$ and $\bar{\phi}_\text{loop}$.  We may think of $\bar{\phi}_\text{laser}$ as a sum of local measurements of the positions of each interferometer arm with respect to the lasers during the atom-light interactions.  The phase $\bar{\phi}_\text{loop}$, on the other hand, represents the phase shift due to the proper time difference\cite{Dimopoulos2008} between the interferometer arms (Fig. 4).

\begin{figure}[h]
	\begin{center}
		\includegraphics{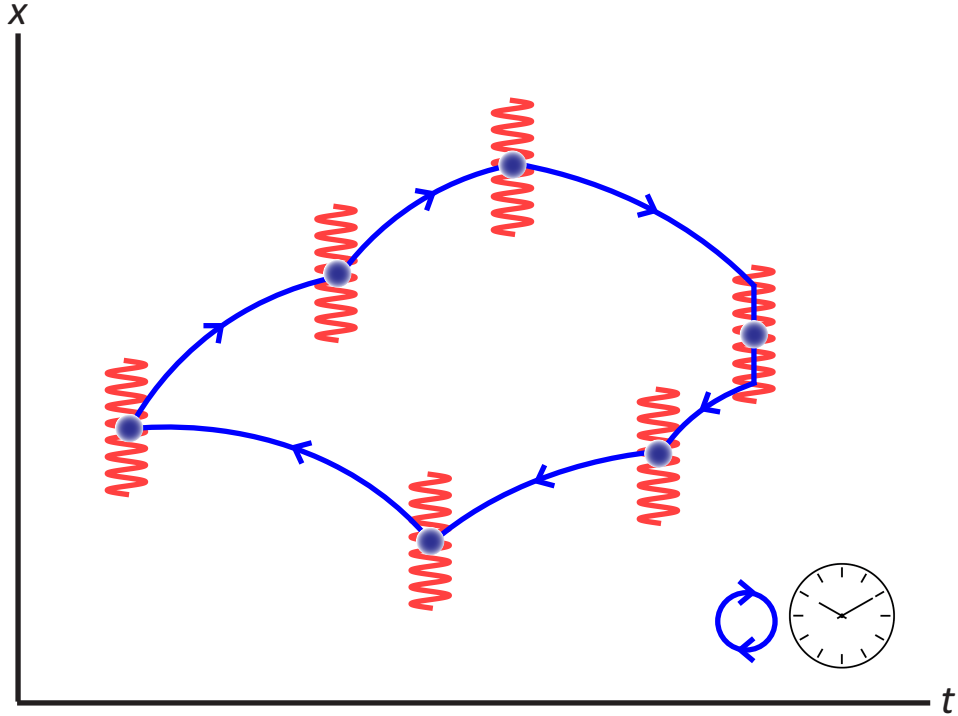}
        \caption{Atom interferometer as described by the semiclassical formalism.  The phase is the sum of two terms.  The first term, $\bar{\phi}_\text{laser}$, is determined by the positions of the arms during the atom-light interactions.  In open interferometers, $\bar{\phi}_\text{laser}$ is computed by using the average position of the two arms during the final beamsplitter.  The second term, $\bar{\phi}_\text{loop}$, is the proper time evolved around the interferometer trajectory (blue).  This term contains the phase $\phi_\text{potential}$, which arises from the potential energy difference between arms.}
	\end{center}
\end{figure}

Further physical insight can be gained by considering the relationship between $\bar{\phi}_\text{laser} + \bar{\phi}_\text{loop}$ and the phase computed from the midpoint theorem.  As we will show in Section \ref{Sec:PropToMP}, whenever the semiclassical approximation is valid, the interferometer phase $\phi$ is given by   
\begin{align}
\phi &= \bar{\phi}_\text{laser} + \bar{\phi}_\text{loop} \\
&= \bar{\phi}_\text{laser} + \sum_i \left(\frac{k_{1,i}+k_{2,i}}{2} \right) [x_2(t_i) - x_1(t_i)] + \phi_\text{potential} \label{Eq:avgLaserNonlocal} \\
&= \phi_\text{MP} + \phi_\text{potential}. \label{Eq:MidpointNonlocal}
\end{align}
Here $\phi_\text{MP}$ is the midpoint phase defined in Eq.~\ref{eq:MPTheorem}, and 
\be
\phi_\text{potential} \equiv  \sum_{n \; \text{odd}} \frac{1}{\hbar} \int_{t_0}^{t_f}  \frac{(n - 1)}{2^{n-1}\;n!} V_n(\bar{x}(t), t) \cdot \Delta x(t)^n \; dt \label{Eq:NonlocalDef}
\ee  
where $V_n$ is the $n^\text{th}$ spatial derivative of the potential energy $V(x,t)$, $\bar{x}(t)$ is the midpoint position of the interferometer at time $t$, and $\Delta x(t) \equiv x_1(t) - x_2(t)$ is the wavepacket separation at time $t$.  The $n = 3$ contribution to $\phi_\text{potential}$ has been derived previously with the Wigner function method,\cite{Dubetsky2016} and an expression equivalent to $\phi_\text{potential}$ has been derived in a representation-free description.\cite{Bertoldi2019}  Note that $\phi_\text{potential}$ can be nonzero without violating the semiclassical approximation, which constrains the high-order spatial dependence of the potential at the scale of the wavepacket size but not at the scale of the wavepacket separation.      

One of the terms from $\bar{\phi}_\text{loop}$ (the second term on the right-hand side of Eq.~\ref{Eq:avgLaserNonlocal}) can be combined with $\bar{\phi}_\text{laser}$ to give $\phi_\text{MP}$.  This term appears in $\bar{\phi}_\text{loop}$ because, in general, an atom-light interaction that changes the interferometer midpoint trajectory induces a proper time difference between the two arms.  The midpoint theorem combines the phase imprinted during the atom-light interaction with the phase due to the light-induced proper time difference, treating them as a single phase that arises from an effective atom-light interaction.  For recoil-insensitive interferometer geometries, e.g. Mach-Zehnder interferometers, $\phi_\text{MP}~=~\bar{\phi}_\text{laser}$. 

We can see from Eq.~\ref{Eq:MidpointNonlocal} and Eq.~\ref{Eq:NonlocalDef} that the midpoint theorem gives the full interferometer phase if $V(x,t)$ is of degree $2$ or lower in $x$.  In such a potential, it is possible to describe the phase evolution entirely in terms of the effective atom-light interactions.  The information needed to compute the phase consists of the laser wavevector and the interferometer midpoint position at the time of each interaction.  The potential energy affects the phase solely by changing the interferometer position with respect to the lasers.

On the other hand, a phase shift beyond the midpoint theorem can occur if the potential energy is of degree greater than $2$ on the length scale of the wavepacket separation.  This effect is represented by $\phi_\text{potential}$.  We may think of $\phi_\text{potential}$ as the part of the proper time difference $\bar{\phi}_\text{loop}$ that cannot be inferred from the atom-light interactions.  This phase is nonlocal in the sense that it cannot be expressed in terms of the interferometer midpoint position, unlike $\phi_\text{MP}$.  Instead, $\phi_\text{potential}$ is sensitive to the potential energy difference over the wavepacket separation.

When $\phi_\text{potential}$ is nonzero, the analogy between the matter-wave interferometer and a classical test object presented in Sections \ref{Sec:classicalAccel} and \ref{Sec:AIAccel} breaks down, and it is no longer sensible to think of the interferometer as performing a set of position measurements.  We discuss the physical implications of $\phi_\text{potential}$ in Section \ref{Sec:Discussion}.

\section{Relationship between the midpoint theorem and the semiclassical method}

\label{Sec:PropToMP}

In this section, we will prove Eq.~\ref{Eq:MidpointNonlocal} and Eq.~\ref{Eq:NonlocalDef}.  Readers who are uninterested in the technical details of the derivation may proceed directly to Section \ref{Sec:Discussion}.

\subsection{Proof}

Defining 
\be
\bar{x}(t) = \frac{x_1(t) + x_2(t)}{2}
\ee
and
\be
\Delta x(t) = x_1(t) - x_2(t),
\ee
the propagation phase is 
\begin{align}
\phi_\text{prop} = \frac{1}{\hbar} \int_{t_0}^{t_f}& \left[\frac{1}{2}\,m\, \dot{x}_1^2 - V(x_1, t) \right]\, dt - \frac{1}{\hbar}\int_{t_0}^{t_f} \left[\frac{1}{2}\,m\, \dot{x}_2^2 - V(x_2, t) \right]\, dt \\
&= \frac{1}{\hbar}\int_{t_0}^{t_f} \left(m\, \Delta \dot{x}\, \dot{\bar{x}} - [V(x_1, t) - V(x_2, t)] \right)\, dt
\end{align}
where a dotted variable represents the time derivative of the variable.  Integrating the first term by parts, we obtain 
\begin{align}
\phi_\text{prop} &= \frac{1}{\hbar}\bigg[m\, \Delta x\, \dot{\bar{x}} \bigg]_{t_0}^{t_f} + \frac{1}{\hbar}\int_{t_0}^{t_f} \left(-m\, \Delta x\, \ddot{\bar{x}} - [V(x_1, t) - V(x_2, t)] \right)\, dt \\
&= \frac{m}{\hbar}\, \Delta x(t_f)\, \dot{\bar{x}}(t_f) + \frac{1}{\hbar}\int_{t_0}^{t_f} \left(-m\, \Delta x\, \ddot{\bar{x}} - [V(x_1, t) - V(x_2, t)] \right)\, dt.
\end{align}
The average separation phase is 
\be
\bar{\phi}_\text{sep} = -\frac{m}{\hbar} \; \Delta x(t_f)\; \dot{\bar{x}}(t_f),
\ee
which cancels with the boundary term in $\phi_\text{prop}$, so we have 
\begin{align} \label{Eq:PhiLoop}
\bar{\phi}_\text{loop} = \phi_\text{prop} + \bar{\phi}_\text{sep} &= \frac{1}{\hbar}\int_{t_0}^{t_f} \left(-m\, \Delta x\, \ddot{\bar{x}} - [V(x_1, t) - V(x_2, t)] \right)\, dt. 
\end{align}

Next, we compute the portion of $\bar{\phi}_\text{loop}$ accumulated between the two-level couplings (i.e. during the drift time).  Using the short-pulse approximation,$^{38}$
we will treat the atom-light interactions as occurring at discrete times $\{t_i\}$ with durations $\{\delta t_i\}$ that are negligible compared to the interferometer time $t_f - t_0$.  Note that guided interferometers can be described in this formalism by including the guiding potential in $V(x,t)$, if the momentum transfer is approximately continuous, or by including additional two-level couplings and associated phase shifts if the momentum transfer is quantized.\cite{Xu2019}  During the drift time, we have 
\be \label{Eq:35}
-m\, \ddot{\bar{x}} = \frac{1}{2} \left(\frac{\partial V(x_1,t)}{\partial x} + \frac{\partial V(x_2,t)}{\partial x}  \right).
\ee
By Taylor expansion of the potential energy around $\bar{x}(t)$, we obtain 
\be
V(x_1, t) = \sum_{n = 0}^\infty \frac{V_n (\bar{x}, t)}{n!} (x_1 - \bar{x})^n = \sum_{n = 0}^\infty \frac{V_n (\bar{x}, t)}{n!} \left(\frac{\Delta x}{2}  \right)^n
\ee
and
\be
V(x_2, t) = \sum_{n = 0}^\infty \frac{V_n (\bar{x}, t)}{n!} (x_2 - \bar{x})^n = \sum_{n = 0}^\infty \frac{V_n (\bar{x}, t)}{n!} (-1)^n \left(\frac{\Delta x}{2}  \right)^n
\ee
where $V_n$ is the $n^\text{th}$ spatial derivative of $V$. Likewise, we have 
\be
\frac{\partial V(x_1,t)}{\partial x} = \sum_{n = 1}^\infty \frac{n\, V_n (\bar{x}, t)}{n!} \left(\frac{\Delta x}{2}  \right)^{n-1}
\ee
and 
\be \label{Eq:39}
\frac{\partial V(x_2,t)}{\partial x} = \sum_{n = 1}^\infty \frac{n\, V_n (\bar{x}, t)}{n!} (-1)^{n-1} \left(\frac{\Delta x}{2}  \right)^{n-1}.
\ee
Using Eqs. \ref{Eq:35} to \ref{Eq:39}, we can rewrite the integrand of Eq. \ref{Eq:PhiLoop} as
\be
\sum_{n\; \text{odd}} \frac{(n - 1)}{2^{n-1}\,n!}\cdot V_n(\bar{x},t)\cdot (\Delta x)^n
\ee
which implies that the portion of $\bar{\phi}_\text{loop}$ accumulated in between light pulses is 
\be
\sum_{n \; \text{odd}} \frac{1}{\hbar} \int_{t_0}^{t_f}  \frac{(n - 1)}{2^{n-1}\;n!}\cdot V_n(\bar{x}, t) \cdot (\Delta x)^n \; dt = \phi_\text{potential}. 
\ee

Finally, we compute the finite contributions to $\bar{\phi}_\text{loop}$ at the times $\{t_i\}$ when the atom-light interactions occur.  At time $t_i$, momentum $\hbar k_{j,i}$ is transferred to the $j^\text{th}$ interferometer arm over a short duration $\delta t_i$.  This gives rise to an additional acceleration 
\be
\ddot{\bar{x}}_i = \frac{\hbar (k_{1,i} + k_{2,i})}{2m}\cdot \frac{1}{\delta t_i}
\ee
and, via Eq.~\ref{Eq:PhiLoop}, an additional contribution to $\bar{\phi}_\text{loop}$ of
\be
\sum_i \frac{1}{\hbar} \int_{t_i - \delta t_i/2}^{t_i + \delta t_i/2} -m\, \Delta x\, \ddot{\bar{x}}_i\, dt = \sum_i \int_{t_i - \delta t_i/2}^{t_i + \delta t_i/2} - \frac{k_{1,i} + k_{2,i}}{2}\cdot \frac{\Delta x}{\delta t_i} \, dt
\ee
\be
= \sum_i \left(\frac{k_{1,i} + k_{2,i}}{2}\right) [x_2(t_i) - x_1(t_i)].
\ee

Altogether, we have 
\begin{align}
\phi &= \bar{\phi}_\text{laser} + \bar{\phi}_\text{loop} \\ 
&= \bar{\phi}_\text{laser} + \sum_i \left(\frac{k_{1,i} + k_{2,i}}{2}\right) [x_2(t_i) - x_1(t_i)] + \phi_\text{potential}.
\end{align}
We conclude by noting that 
\be
\bar{\phi}_\text{laser} + \sum_i \left(\frac{k_{1,i} + k_{2,i}}{2}\right) [x_2(t_i) - x_1(t_i)]  = \phi_\text{MP}
\ee
where $\phi_\text{MP}$ is the midpoint phase defined in Eq.~\ref{eq:MPTheorem}.  Thus, we have
\be
\phi = \phi_\text{MP} + \phi_\text{potential}
\ee
as desired.

\subsection{Closed-form expression for $\phi_\text{potential}$}

We can obtain a closed-form expression for $\phi_\text{potential}$ from the fact that
\be
-m\, \ddot{\bar{x}} = \frac{1}{2} \left(\frac{\partial V(x_1,t)}{\partial x} + \frac{\partial V(x_2,t)}{\partial x}  \right)
\ee
when the light is off. Eq.~\ref{Eq:PhiLoop} then implies that
\be \label{Eq:phiPotClosed}
\phi_\text{potential} = \frac{1}{\hbar}\int_{t_0}^{t_f} \left(\frac{\Delta x}{2} \left[\frac{\partial V(x_1,t)}{\partial x} + \frac{\partial V(x_2,t)}{\partial x}  \right] - [V(x_1, t) - V(x_2, t)] \right)\; dt. 
\ee
This expression demonstrates that $\phi_\text{potential}$ measures the difference between two quantities:  (i) the potential energy difference between the arms, and (ii) the potential energy difference that one would infer from the average acceleration. 

\section{Discussion} 

\label{Sec:Discussion}

To what extent does a matter-wave interferometer provide the same physical information as a classical measurement?  The answer depends crucially on the spatial structure of the potential energy at the length scale of the wavepacket separation (Fig. 5).  

As we saw in Section \ref{Sec:classicalAccel} and Section \ref{Sec:AIAccel}, a matter-wave interferometer is analogous to a classical measuring device if the potential energy can be well-approximated by a function of degree 2 or lower in position coordinates.  In this case, the phase response is determined by the interferometer's midpoint positions at the times of the beamsplitter and mirror interactions.  The ``ruler'' that measures these positions is the wavevector associated with the momentum transfer of each interaction.  Dependence of the phase on other parameters, such as mass or charge, can arise only if the experiment is designed in such a way that the momentum transfer or midpoint position is a function of those parameters (as is the case in recoil interferometers\cite{Borde1989}).  Regardless of the experimental design, if the interferometer phase is rewritten in terms of the momentum transfer and the midpoint trajectory, all other parametric dependence vanishes.  The phase does not depend on the thermal de Broglie wavelength\cite{Cronin2009} $\sqrt{2\pi \hbar^2 / (m k_B T)}$ or on the Compton frequency\cite{Lan2013} $m c^2/\hbar$.  

In the context of gravity, the absence of mass-dependent phase shifts in a uniform field is an illustration of the equivalence principle, which states that the trajectory of a test particle in a uniform gravitational field is independent of its mass and composition.\cite{DiCasola2015}  Matter-wave interferometers in a gravitational potential of degree 1 can test the equivalence principle but cannot provide any further information about the gravitational interaction.  A recent observation of an atom interferometer evolving in a region with nontrivial spacetime curvature\cite{Asenbaum2017} demonstrated that the midpoint theorem remains experimentally valid for a gravitational potential of degree 2.  The interferometer phase is determined by the acceleration of the midpoint trajectory, which in that case is resolvably different from the acceleration of either interferometer arm.            

On the other hand, as shown in Section \ref{Sec:BeyondMP} and Section \ref{Sec:PropToMP}, the interferometer phase obtains an additional contribution $\phi_\text{potential}$ if the potential energy is of degree 3 or higher.  This phase shift depends on the potential energy difference between interferometer arms, and it is nonzero if the potential energy contains high-order terms that are asymmetric around the midpoint trajectory.  Unlike the midpoint phase, $\phi_\text{potential}$ cannot be written in terms of the midpoint trajectory and can depend nontrivially on mass or charge. Nevertheless, the phase remains independent of the thermal de Broglie wavelength and the Compton frequency; the relevant energy scale for computing $\phi_\text{potential}$ is the potential energy difference between arms. 

\begin{figure}[t]
	\begin{center}
		\includegraphics[width=\columnwidth]{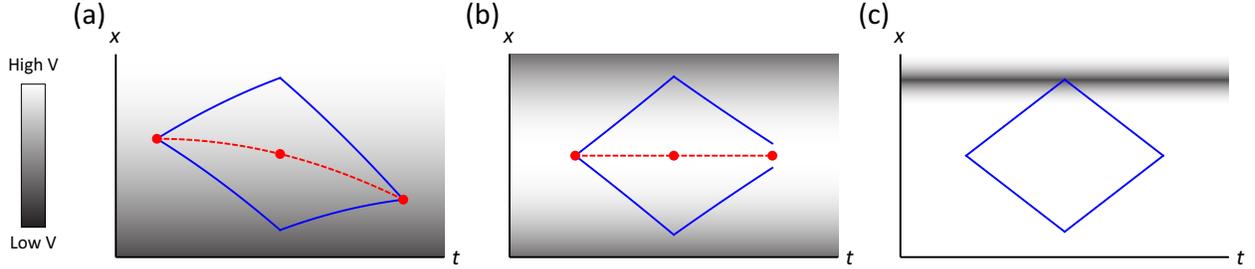}
        \caption{Dependence of interferometer behavior on spatial structure of potential energy $V$ (magnitude indicated with gray shading). (a) Linear potential; midpoint theorem applies.  The midpoint trajectory (red) has the same acceleration as each of the interferometer arms (blue).  (b) Quadratic potential; midpoint theorem applies.  The acceleration of the midpoint trajectory differs from that of either arm.\cite{Asenbaum2017} (c) High-order potential; beyond the midpoint theorem.  The phase is decoupled from the midpoint trajectory and depends on the potential energy difference between arms.}
		\label{fig:figure1}
	\end{center}
\end{figure}

The phase shift $\phi_\text{potential}$ decouples the interferometer phase from the arm trajectories.  By choosing an appropriate potential, one can realize a situation in which the arms are deflected by the potential, yet there is no interferometer phase response because $\phi_\text{potential}$ cancels the midpoint phase.  Even more surprisingly, there exist potentials in which the interferometer phase is nonzero even though the potential does not induce any deflections at all.  This can be seen by considering a time-dependent potential, where $V(x_1, t) - V(x_2, t)$ may be nonzero even if $\frac{\partial V(x_1, t)}{\partial x} = \frac{\partial V(x_2, t)}{\partial x} = 0$ for all $t$.  An example of such a potential is presented in Appendix A.  

Since $\phi_\text{potential}$ can be nonzero even in the absence of trajectory perturbations, it is clear that $\phi_\text{potential}$ contains information about the potential energy beyond that of any classical position measurements.  Physically, $\phi_\text{potential}$ describes the proper time difference between arms that is induced by the potential.  The observation of the electromagnetic Aharonov-Bohm effect\cite{Aharonov1959,Chambers1960,Werner2010} is a well-known example of a matter-wave interferometry experiment in which $\phi_\text{potential}$ is nonzero.  

Notably, however, no experiment has yet observed a nonzero $\phi_\text{potential}$ induced by a gravitational potential.  The observation of such a ``gravitational Aharonov-Bohm effect''\cite{Hohensee2012} would demonstrate unambiguously that a matter-wave interferometer is sensitive to gravitationally induced proper time differences between arms.  Such a measurement would also be the first observation of gravitational time dilation in a single quantum system. 

\subsection*{Acknowledgments}

We thank Tim Kovachy for fruitful discussions.  This work was supported by the Vannevar Bush Faculty Fellowship Program.

\appendix 

\section{Example calculation of the phase shift in a high-order potential}

Suppose a Mach-Zehnder matter-wave interferometer with beamsplitter wavevector $k$ and mass $m$ evolves with potential energy $V = 0$ and drift time $T$ between pulses.  Choosing coordinates so that the interferometer is symmetric around the origin, the arm trajectories are given by 
\be
x_1(t) = \begin{cases} \frac{\hbar k}{2 m} t,  & 0\le t < T\\
-\frac{\hbar k}{2 m} (t - 2T), & T< t \le 2T
\end{cases}
\ee
and 
\be
x_2(t) = \begin{cases} -\frac{\hbar k}{2 m} t,  & 0\le t < T\\
\frac{\hbar k}{2 m} (t - 2T), & T< t \le 2T.
\end{cases}
\ee
The interferometer phase $\phi$ is given by 
\be
\phi = \phi_\text{MP} = 0.
\ee
Now suppose that the same interferometer evolves with potential energy
\be
V(x, t) = \begin{cases} A \left[ \frac{1}{3}x^3 - \left(\frac{\hbar k}{2 m} t\right)^2 x \right],  & 0\le t < T\\
A \left[ \frac{1}{3}x^3 - \left(\frac{\hbar k}{2 m} (t-2T)\right)^2 x \right], & T< t \le 2T.
\end{cases} 
\ee
This function has the property that $\frac{\partial V(x_1)}{\partial x} = \frac{\partial V(x_2)}{\partial x} = 0$ for $0 < t < 2T$, so the potential energy does not alter the arm trajectories, and $\phi_\text{MP} = 0$.  Nevertheless, the potential energy induces a phase shift
\be
\phi = \phi_\text{potential} = \frac{\hbar^2\, k^3\, A\, T^4}{12\, m^3}.
\ee
This example demonstrates that in a high-order potential, the interferometer phase is decoupled from the arm trajectories and can be nonzero even if the potential energy does not cause any deflections.  

\section{Perturbative semiclassical methods}

In the semiclassical method presented in Section \ref{Sec:BeyondMP} and Section \ref{Sec:PropToMP}, the action difference between arms is calculated self-consistently by integrating the Lagrangian along arm trajectories that are solutions of the equations of motion.  It is also possible to compute the interferometer phase perturbatively by treating part of the potential energy as a perturbation $\delta V$ and integrating $\delta V$ along the unperturbed arm trajectories.  A number of authors\cite{Anandan1984,Werner1994} have used this technique to calculate low-order phase shifts of a matter-wave interferometer in various potentials.  

While the perturbative approach is a mathematically sound way to compute the interferometer phase,\cite{Storey1994} one must be careful not to make incorrect assertions about the action difference between arms on the basis of a perturbative calculation.  In a potential of degree 2 or lower, the action of a trajectory that solves the equations of motion depends only on the positions and momenta of its endpoints.\cite{Antoine2003}  In particular, a Mach-Zehnder interferometer in such a potential has zero action difference between arms.  The $kgT^2$ phase shift of a Mach-Zehnder interferometer in a uniform gravitational field is not due to any action difference but rather due to the relative acceleration between the freely falling particles and the accelerated beamsplitters and mirrors.\cite{Wolf2011} 

\section{State-dependent potentials}

Our presentation of the semiclassical method assumes that the potential energy is solely a function of position and time.  It is also possible to use the semiclassical formalism to compute the phase of an interferometer in which the potential energy depends on an additional state label $l$.  In that case, the two arms evolve in distinct potentials whenever the internal states have different values of $l$.  

One proposal\cite{Zimmermann2017} for such an interferometer suggests interfering distinct magnetically sensitive states in a uniform magnetic gradient.  In this proposal, $l$ corresponds to the magnetic dipole moment of the internal state, which differs between interferometer arms.  Since the potential energy is a nontrivial function of $l$, the interferometer phase is not given by Eq.~\ref{eq:MPTheorem} but by Eq.~\ref{Eq:MidpointNonlocal}, using the closed-form expression in Eq.~\ref{Eq:phiPotClosed} for $\phi_\text{potential}$ and including the dependence of the potential energy on $l$.  

Since the potential energy of each internal state in this example depends linearly on position, it remains true that the action of each trajectory segment depends only on the positions and momenta of its endpoints.\cite{Antoine2003}  From the perspective of the midpoint theorem, the relevance of a low-order state-dependent potential is that state dependence can provide an additional source of displacement between interferometer arms.  One could observe a scalar Aharonov-Bohm effect in a high-order state-dependent potential or in a state-dependent potential with nontrivial time dependence.\cite{Roura2014}


%

\end{document}